\documentstyle[aps,12pt]{revtex}
\def\Journal#1#2#3#4{{#1} {#2} (#4) #3}

\def\NPA{{\rm Nucl. Phys.} A}
\def\NPB{{\rm Nucl. Phys.} B}
\def\PLB{{\rm Phys. Lett.}  B}
\def\PRL{\rm Phys. Rev. Lett.}
\def\PRD{{\rm Phys. Rev.} D}

\def\al{\alpha}

\def\be{\begin{equation}}
\def\ee{\end{equation}}
\def\bea{\begin{eqnarray}}
\def\eea{\end{eqnarray}}
\input{psfig.sty}
\begin{document}
\vspace{2.in}
\title{ Exploring the timelike region for the elastic form factor in  
the light-front quantization}
\author{ Ho-Meoyng Choi and Chueng-Ryong Ji }
\address{ Department of Physics,
North Carolina State University,
Raleigh, N.C. 27695-8202}
\maketitle

\narrowtext
\vspace{1.0in}
\begin{abstract}
Even though the Drell-Yan-West formulation is the most rigorous and 
well-established framework to compute the exclusive processes, its utility 
has been limited only to
the spacelike region because of the intrinsic kinematic constraint $q^+=0$.
We present an explicit example demonstrating how one may obtain the necessary
information (i.e. nonvalence or so called Z-graph contribution) in the timelike 
region of exclusive process without encountering a formidable task of direct 
calculation that has hindered so far the progress in this area. 
In the analysis of $q\bar{Q}$ bound state form factors using
an exactly solvable model of $(3+1)$ dimensional scalar field theory 
interacting with gauge fields, the results
analytically continued from the spacelike region coincide exactly
with the direct results in the timelike region. This example verifies 
that the method of analytic continuation is capable of yielding 
the effect of complicate nonvalence contributions.
The meson peaks analogous to the vector meson dominance(VMD) phenomena
are also generated at the usual VMD positions.
\end{abstract}

PACS: 11.15.Bt, 11.10.Kk, 11.40.-q

Keywords: Time-like form factor, Analytic continuation, Dispersion relation 
\newpage 
\section{ Introduction} 
The Drell-Yan-West ($q^{+}=q^{0}+q^{3}=0$) frame in the light-front
quantization provided an effective formulation for the calculation of various 
form factors in the spacelike momentum transfer region 
$q^{2}=-Q^{2}<0$\cite{LB}. As an example, only parton-number-conserving Fock 
state (valence) contribution (e.g. Fig. 1(a))
is needed in $q^+=0$ frame when the ``good" 
components of the current, $j^{+}$ and $j_{\perp}=(j_{x},j_{y})$, are 
used\cite{Mel1} for the spacelike electromagnetic form factor 
calculation of pseudoscalar mesons. Successful light-front quark 
model (LFQM) description of various pseudoscalar form factors 
can also be found in the literatures\cite{CJ2,Chung,Jaus,Card1,CJ1}.
However, not all is well. For the higher spin system  composed of fermions,
one needs in general to consider the nonvalence contribution (e.g. Fig.~1(b)) 
to fulfill the covariance requirement even when $j^+$ and $j_{\perp}$ 
are used in the $q^+=0$ frame. The nonvalence contribution in the $q^+=0$ frame
is called the ``zero-mode". The details of zero-mode contributions for the 
spin-one system with the $j^+$ current coupled with longitudinal polarization 
vector as well as the semileptonic transition form factor of pseudoscalar
meson with the $j_{\perp}$ current can be seen in 
Refs.~\cite{Keister,Mel2,Ja1}.

Furthermore, the timelike ($q^{2}>0$) form factor analysis in the light-front
quark model has been hindered by the fact that $q^{+}=0$ frame is defined
only in the spacelike region ($q^{2}=q^{+}q^{-}-q^{2}_{\perp}<0$). 
While the $q^{+}\neq0$ frame can be used in principle to compute the timelike
form factors, it is inevitable (if $q^{+}\neq 0$) to encounter the nonvalence
diagram arising from the quark-antiquark pair creation (so called ``Z-graph") 
as shown in Fig. 1(b). 
The main source of difficulty in calculating the nonvalence 
diagram (Fig. 1(b)) is the lack of information on the black blob which
should contrast with the white blob representing the usual light-front
valence wave function. In fact, it was reported\cite{CJ2} that
the omission of nonvalence contribution
leads to a large deviation from the full results. 
The timelike form factors associated with the hadron pair productions in
$e^{+}e^{-}$ annihilations also involve the nonvalence contributions. 
Therefore, it is not a simple task to explore the timelike region with the
framework at hand and the investigation is still just at the beginning stage
to the best of our knowledge.
\begin{figure}
\psfig{figure=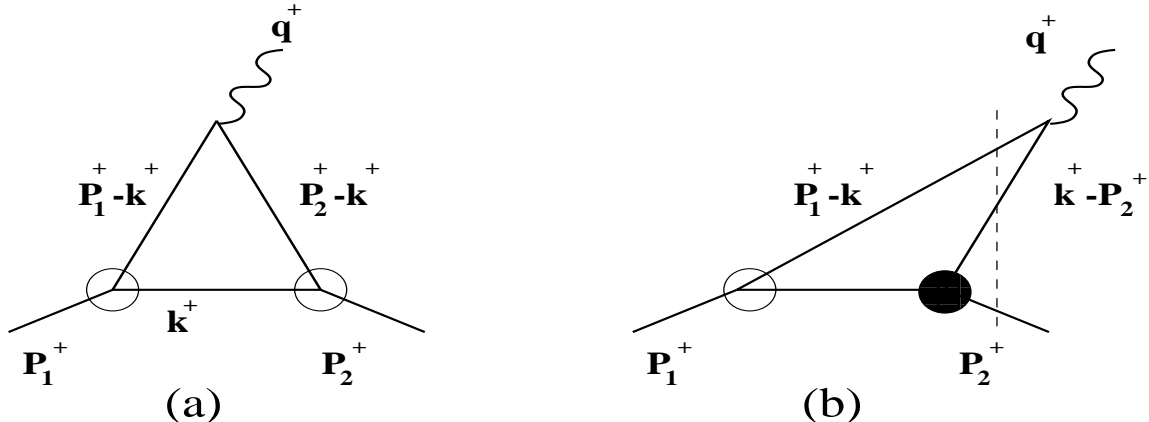,width=6in,height=2.2in}
\caption{The light-front quark model description of a electroweak
meson form factor:
(a) the usual light-front valence diagram and (b) the
nonvalence(pair-creation) diagram. The vertical dashed line in (b)
indicates the energy-denominator for the nonvalence contributions.
While the white blob represents the usual light-front
valence wave function, the modeling of black blob has not yet been made.}
\end{figure}
The aim of this paper is to provide at least a clear example of demonstration
showing how one might take advantage of the existing formulation to get the new
information (i.e. nonvalence contribution or so called Z-graph) necessary 
in the timelike region of exclusive processes. 
This can be done by the analytic continuation from the spacelike form 
factor calculated in the Drell-Yan-West ($q^{+}=0$) frame to the timelike
region. In this paper, we focus on the case that does not require the zero-mode
contributions even though one can in principle apply the same method 
presented in this work to the case of zero-mode contributions once
they are identified. However, in the case of the 
zero-mode contributions, an additional step is necessary to identify  
them\footnote{An example including the zero-mode 
contribution can be shown in the calculation of the $f_-$ form factor 
for the exclusive semileptonic decays between two pseudoscalar mesons 
using the fermionic covariant model with the monopole smearing 
vertices~\cite{Mel1,Mel2}. In this case, we indeed verified the applicability 
of our method by checking if the identical results are obtained when different 
components of the current are used in different frames, i.e., $q^+\neq0$ 
and $q^+=0$ frames. See also the discussion in Section V.}.

For an explicit demonstration of our work, we use an exactly solvable 
model of ($3+1$) dimensional scalar field theory interacting with gauge 
fields. The Lagrangian of the system is given by~\cite{BH}
\bea\label{eq:Lag}
{\cal L}&=& (\partial_\mu\phi_a + ie_a A_\mu\phi_a)^{\dagger}
(\partial^\mu\phi_a + ie_a A^\mu\phi_a) - m^{2}_{a}\phi^{\dagger}_{a}\phi_a
\nonumber\\
&&+ (\partial_\mu\phi_b + ie_b A_\mu\phi_b)^{\dagger}
(\partial^\mu\phi_b + ie_b A^\mu\phi_b) - m^{2}_{b}\phi^{\dagger}_{b}\phi_b
\nonumber\\
&&-\frac{1}{2}(M^2\Phi^2 
- \partial_{\mu}\Phi\partial^{\mu}\Phi)
+ g\Phi(\phi^{\dagger}_a\phi_b + \phi^{\dagger}_b\phi_a),
\eea
where $\phi_{a(b)}$ corresponds to the bosonic quark field with the mass 
$m_{a(b)}$ and $\Phi$ is the bound state meson with the mass $M$.
Our model is essentially the $(3+1)$ dimensional extension of Mankiewicz
and Sawicki's $(1+1)$ dimensional quantum field theory model \cite{SM}, 
which was later reinvestigated by several 
others \cite{BH,GS,Sa,Choi,Ba}.
The starting model wave function is the \underline{solution} of covariant
Bethe-Salpeter (BS) equation in the ladder approximation with a
relativistic version of the contact interaction \cite{SM}.
Here, we do not take the Hamiltonian approach. The covariant model wave
function is a product of two free single particle propagators, the overall
momentum-conservation Dirac delta, and a constant vertex function.
Consequently, all our form factor calculations are nothing
but various ways of evaluating the Feynman perturbation-theory triangle
diagram in scalar field theory.

Even though the model may not be as realistic as one may 
want, it achieves the goal of being the first example up to our knowledge. 
We think that it provides at least some guidance how one can persue the 
same (or similar) idea in the more realistic (perhaps more phenomenological 
in that respect) models. In this model, we calculate: 
(A) the timelike process of $\gamma^{*}\to M + \bar{M}$ 
transition in $q^{+}\neq 0$ ($q^{2}>0$) frame, 
(B) the spacelike process of $M\to\gamma^{*} +M$ in $q^{+}\neq0$ ($q^{2}<0$) 
frame, and   
(C) the spacelike process of $M\to\gamma^{*} +M$ in $q^{+}=0$ frame.  
Using the analytic continuation from $q^{2}<0$ to $q^{2}>0$, we
show that the result in (C), which is obtained without encountering the 
nonvalence contributions at all, exactly reproduces the result in (A). 
In fact, all three results (A), (B), and (C) coincide with each other in the
entire $q^{2}$ range.
We also confirm that our results are consistent with the dispersion 
relations\cite{Drell,Gasio,Man,Gell}. We consider not only for the equal 
quark/antiquark mass case such as the pion but also for the unequal
mass cases such as $K$ and $D$.

The paper is organized as follows: In Sec. II, we derive the timelike 
electromagnetic (EM) form factor of $\gamma^{*}\to M+\bar{M}$ process in the 
$q^{+}\neq 0$ frame (A) 
and discuss the singularities occuring from the on-energy shell 
of quark-antiquark pair creation.  
In Sec. III, the spacelike form factor of $M\to\gamma^{*} + M$ process 
is calculated both in the $q^{+}\neq0$ (B) and $q^{+}=0$ (C) frames. 
We then analytically continue the spacelike form factors to the timelike 
region. The singularities occured in the timelike region are also discussed. 
In Sec. IV, for the numerical calculation of the 
EM ($\pi$, $K$, and $D$) meson 
form factors for three different cases (A), (B)
and (C), we use the constituent quark and antiquark masses
($m_{u}=m_{d}=0.25$ GeV, 
$m_{s}=0.48$ GeV, and $m_{c}=1.8$ GeV)\cite{CJ1,CJ2,CJ3}
and show that the form factors obtained from those three
different cases are indeed equal to each other for the entire $q^{2}$
region. 
The meson peaks analogous to the vector meson dominance(VMD) are also
obtained.
The conclusion and discussion follows in Sec. V.  
\section{ Form factors in the timelike region } 
The EM local current $J^{\mu}(0)$
responsible for a virtual photon decay into two $q\bar{Q}$ bound states 
in the scalar field theory can be calculated using the diagrams shown in
Fig. 2. 
The covariant diagram shown in Fig. 2(a) is equivalent to the sum of 
two light-front time-ordered diagrams in Figs. 2(b) and 2(c).  
The EM current $J^{\mu}(0)$ obtained from the covariant 
diagram of Fig. 2(a) is given by
\bea\label{eq:EMCOV}
J^{\mu}(0)&=& i e_{q}g^2\int d^{4}k\frac{1}{ (q-k)^{2}-m_{q}^{2}+i\epsilon}
(q-2k)^{\mu}\frac{1}{ (q-k-P_{2})^{2}-m_{\bar{Q}}^{2}+i\epsilon}
\nonumber\\
&&\times\frac{1}{ k^{2}-m_{q}^{2}+i\epsilon}\;\; 
+\;\; e_{\bar{Q}}(m_{q}\leftrightarrow m_{\bar{Q}}\; {\rm of\; the\; 
first\; term}), 
\eea
where $m_{q(\bar{Q})}$ and $e_{q(\bar{Q})}$ are the constituent 
quark (antiquark) mass and charge, respectively. 
\begin{figure}
\psfig{figure=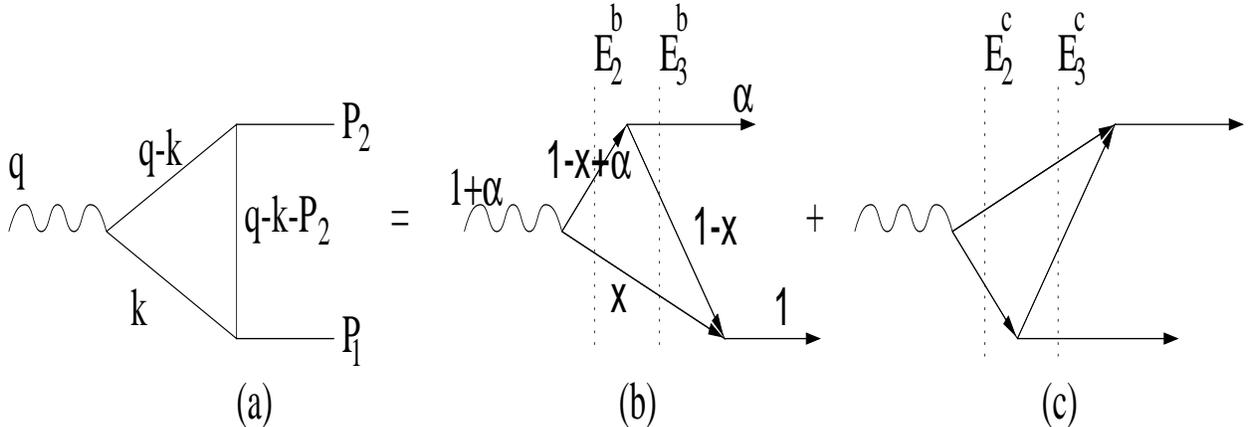,width=6.5in,height=2.2in}
\caption{The electromagnetic decays of a photon into two-body bound
sates, i.e., $\gamma^{*}\to q\bar{q}$(or $Q\bar{Q}$)$\to
{\cal M}(q\bar{Q}){\cal M}(q\bar{Q})$,
in scalar theory: Covariant representation (a),
and the light-front time ordered contributions to the decay amplitude (b)
and (c).}
\end{figure}
The corresponding form factor $F(q^2)$ of the $q{\bar{Q}}$ bound state in
timelike($q^{2}>0$) region is defined by 
\bea\label{eq:Jtime}
J^{\mu}(0)= (P_{1}-P_{2})^{\mu}F(q^{2}), 
\eea
where $q=P_{1}+P_{2}$, $P^{2}_{1}=P^{2}_{2}=M^{2}$ and $M$ is the 
mass of a $q\bar{Q}$ bound state scalar particle. 

Using the Cauchy integration over $k^-$ in Eq.~(\ref{eq:EMCOV}), 
we can find each
time-ordered contribution (Figs. 2(b) and 2(c)) to the timelike
form factor $F(q^2 > 4M^2)$ in Eq.~(\ref{eq:Jtime}).
This procedure allows us to analyze the singularity structure
of each light-front time-ordered diagram as well.
For the calculation of each light-front time-ordered contribution,
we take the purely longitudinal momentum frame, i.e.,  
$q^{+}\neq 0$,  $q_{\perp}=0$ and $P_{1\perp}=P_{2\perp}=0$.
Accordingly, $q^{2}(=q^{+}q^{-}) > 4 M^2$ is given by  
\bea\label{eq:q2time} 
q^{2}=M^{2}(1+\al)^{2}/\al,
\eea
where $\al=P^{+}_{2}/P^{+}_{1}=q^{+}/P^{+}_{1}-1$ is 
the longitudinal momentum fraction and 
the two solutions for $\al$ are given by
\bea\label{eq:alpha}
\al_{\pm}=\biggl(\frac{q^{2}}{2M^{2}}-1\biggr)\pm
\sqrt{\biggl(\frac{q^{2}}{2M^{2}}-1\biggr)^{2}-1}.
\eea   
Note that both $\al_{\pm}=1$ correspond to the threshold, 
$q^{2}=4M^{2}$. The EM form factor $F(q^{2})$ in Eq.~(\ref{eq:Jtime})
is independent of the subscript sign of $\al$. Thus, 
one can take either ${\al}_{+}$ or ${\al}_{-}$ to calculate $F(q^2)$.
Here, for convenience, we use 
$\al=\al_{-}$ which ranges from 0 to 1 
for the physical momentum transfer region, i.e., $\al_{-}\to0$ as 
$q^{2}\to\infty$ and 1 as $q^{2}\to 4M^2$. Of course, we can use
$\al=\al_{+}$ equally well and verify that the two results (${\al}_{+}$
and ${\al}_{-}$) are exactly same
for the calculation of $F(q^{2})$.
 
Since $q^{+}>P^{+}_{1}\geq P^{+}_{2}>0$ for $\al=\al_{-}$, 
the Cauchy integration over $k^{-}$ in Eq.~(\ref{eq:EMCOV}) has 
two nonzero contributions to the residue calculations, one coming from
the interval (i) $0<k^{+}<P^{+}_{1}$ (see Fig. 2(b)) and the other 
from (ii) $P^{+}_{1}<k^{+}<q^{+}$ (see Fig. 2(c)). The internal
momentum $k^{+}$ is defined by $k^{+}=xP^{+}_{1}$, where $x$ 
is the Lorentz invariant longitudinal momentum variable. 
The ``good" component of the current, $J^{+}(0)$, is used in our computation 
of the two light-front diagrams Figs. 2(b) and 2(c).
In the following, for simplicity, we will explicitly write
neither the obvious second term in Eq.~(\ref{eq:EMCOV}) nor the charge factor
($e_q$ or $e_{\bar{Q}}$).

In the region of $0<k^{+}<P^{+}_{1}$, the residue is at the 
pole of $k^{-}=[m_{q}^{2}+k^{2}_{\perp}-i\epsilon]/k^{+}$, which is placed 
in the lower half of complex-$k^{-}$ plane. Thus, the Cauchy integration of 
$J^{+}$ in Eq.~(\ref{eq:EMCOV}) over $k^{-}$ in this region gives   
\bea\label{eq:JbP} 
J^{+}_{b}(0)&=&\pi g^2\int^{P^{+}_{1}}_{0}
dk^{+}d^{2}k_{\perp}\frac{q^{+}-2k^{+}}
{k^{+}(q^{+}-k^{+})(P^{+}_{1}-k^{+})}\nonumber\\
&&\times \frac{1}{\biggl[ q^{-}-(m_{q}^{2}+k^{2}_{\perp})/k^{+} -
(m_{q}^{2}+k_{\perp}^{2})/(q^{+}-k^{+})\biggr]}\nonumber\\ 
&&\times
\frac{1}{\biggl[ q^{-}-P^{-}_{2}- (m_{q}^{2}+k^{2}_{\perp})/k^{+} -
(m_{\bar{Q}}^{2}+k_{\perp}^{2})/(P_{1}^{+}-k^{+})\biggr]},
\eea 
where the subscript $b$ in $J^{+}_{b}(0)$ implies the current in Fig. 2(b)
and $g$ is fixed by the normalization of the EM form factor at $q^2$=0. 
The last two terms in Eq.~(\ref{eq:JbP}) correspond to the two and three 
particle intermediate states.
We represent the
energy denominators of
the two and three particle intermediate states as $E^{b}_{2}$ and
$E^{b}_{3}$ 
in Fig. 2(b), respectively.  

To analyze the singularities of Eq.~(\ref{eq:JbP}), we further integrate 
over $k_{\perp}$ and obtain  
\bea\label{eq:JbP1}
J^{+}_{b}(0)=\pi^2 g^2\int^{1}_{0}dx
\frac{x(1+\al-2x)/(1+\al)}{{\cal E}^{b}_{3}-{\cal E}^{b}_{2}}
\ln\biggl(\frac{{\cal E}^{b}_{2}}{{\cal E}^{b}_{3}}\biggr),
\eea
where ${\cal E}^{b}_{2}=x(1+\al-x)M^{2}/\al- m_{q}^{2}$ and
${\cal E}^{b}_{3}=x(1-x)M^{2}-[xm_{q}^{2} + (1-x)m_{\bar{Q}}^{2}]$. 
While ${\cal E}^{b}_{3}$ is not zero (${\cal E}^{b}_{3}\neq 0$) in general  
for the entire physical region, ${\cal E}^{b}_{2}$ can be zero 
when $q^{2}\geq4m^{2}_{q(\bar{Q})}$.
The singular structure of ${\cal E}^{b}_{3}-{\cal E}^{b}_{2}$ term in 
Eq.~(\ref{eq:JbP1}) depends on whether a $q\bar{Q}$ bound state scalar 
particle is strongly bounded ($M^{2} < m^{2}_{q} + m^{2}_{\bar{Q}}$) or
weakly bounded ($M^{2} > m^{2}_{q} + m^{2}_{\bar{Q}}$).
As we will show in our numerical calculations (Section IV),
anomalous threshold appears for $M^{2} > m^{2}_{q} + m^{2}_{\bar{Q}}$
while only the normal threshold of bound state exists for
$M^{2} < m^{2}_{q} + m^{2}_{\bar{Q}}$. 

In the region of $P^{+}_{1}<k^{+}<q^{+}$, 
the residue is at the pole of
$k^{-}=q^{-}-[m_{q}^{2}+(q_{\perp}-k_{\perp})^{2}-i\epsilon]/(q^{+}-k^{+})$, 
which is placed in the upper half of complex-$k^{-}$ plane. 
Thus, the Cauchy integration of
$J^{+}(0)$ in Eq.~(\ref{eq:EMCOV}) over $k^{-}$ in this region yields 
the result
\bea\label{eq:JcP}
J^{+}_{c}(0)&=& -\pi g^2\int^{q^{+}}_{P^{+}_{1}}
dk^{+}d^{2}k_{\perp}\frac{q^{+}-2k^{+}}
{k^{+}(q^{+}-k^{+})(P^{+}_{1}-k^{+})}\nonumber\\
&&\times 
\frac{1}{\biggl[ q^{-}- (m_{q}^{2}+k^{2}_{\perp})/k^{+} -
(m_{q}^{2}+k_{\perp}^{2})/(q^{+}-k^{+})\biggr]}\nonumber\\ 
&&\times
\frac{1}{\biggl[ q^{-}-P^{-}_{1} 
+ (m_{\bar{Q}}^{2}+k^{2}_{\perp})/(P^{+}_{1}-k^{+})
- (m_{q}^{2}+k_{\perp}^{2})/(q^{+}-k^{+})\biggr]},
\eea
where the subscript $c$ in $J^{+}_{c}(0)$ means the current in Fig. 2(c). 
After the integration over the $k_{\perp}$ in Eq.~(\ref{eq:JcP}), 
we obtain    
\bea\label{eq:JcP1}
J^{+}_{c}(0)= -\pi^2 g^2\int^{1}_{0} dX 
\frac{\al X(1+\al-2\al X)/(1+\al)}{{\cal E}^{c}_{3}-{\cal E}^{c}_{2}}
\ln\biggl( \frac{ {\cal E}^{c}_{2}}{{\cal E}^{c}_{3}}\biggr),
\eea 
where $x=1+\al (1-X)$, ${\cal E}^{c}_{3}=X(1-X)M^{2}-[X m_{\bar{Q}}^{2} 
+ (1-X)m_{q}^{2}]$ and ${\cal E}^{c}_{2}=\al X(1+ \al -\al X)M^{2}/\al
- m_{q}^{2}$. The pole structure in Eq.~(\ref{eq:JcP1}) is equivalent 
to that of Eq.~(\ref{eq:JbP1}).

Consequently, the timelike form factor in Eq.~(\ref{eq:Jtime}) 
is given by   
\bea\label{eq:EMtime}
F(q^{2}) &=& \frac{\pi^2 g^2}{\al^{2}-1}\int^{1}_{0}dx
\biggl\{ \frac{x(1+\al-2x)}{{\cal E}^{b}_{3}-{\cal E}^{b}_{2}}
\ln\biggl(\frac{{\cal E}^{b}_{2}}{{\cal E}^{b}_{3}}\biggr) 
\nonumber\\
&&- \frac{\al x(1+\al-2\al x)}{{\cal E}^{c}_{3}-{\cal E}^{c}_{2}}
\ln\biggl( \frac{ {\cal E}^{c}_{2}}{{\cal E}^{c}_{3}}\biggr)\biggr\},
\eea
where ${\cal E}^{b}_{2}, {\cal E}^{b}_{3}, {\cal E}^{c}_{2}$ and 
${\cal E}^{c}_{3}$ are defined in Eqs.~(\ref{eq:JbP1}) 
and~(\ref{eq:JcP1}).
\section{Form factors in spacelike region and the analytic continuation 
to the timelike region} 
In this section, we calculate the EM form factor in spacelike momentum
transfer region and then analytically continue to the timelike region 
to compare the result with the timelike form factor 
(i.e. Eq.~(\ref{eq:EMtime})) that we obtained in the previous section.
The EM current of a $q\bar{Q}$ bound state in spacelike 
momentum transfer region is defined by the local current $j^{\mu}(0)$;
\bea\label{eq:Jspace}
j^{\mu}(0)= (P_{1}+P_{2})^{\mu}{\cal F}(q^{2}),
\eea
where $q=P_{1}-P_{2}$, $q^{2}<0$ and ${\cal F}(q^2)$ is the spacelike
form factor. 
The EM current $j^{\mu}(0)$ obtained 
from the covariant triangle diagram of Fig. 3(a) is given by  
\bea\label{eq:JLF}  
j^{\mu}(0)= ig^2\int d^{4}k\frac{1}{ (P_{1}-k)^{2}-m_{q}^{2}+i\epsilon}
(P_{1}+P_{2}-2k)^{\mu}\frac{1}{ (P_{2}-k)^{2}-m_{q}^{2}+i\epsilon}
\frac{1}{ k^{2}-m_{\bar{Q}}^{2}+i\epsilon}.
\eea
\begin{figure}
\psfig{figure=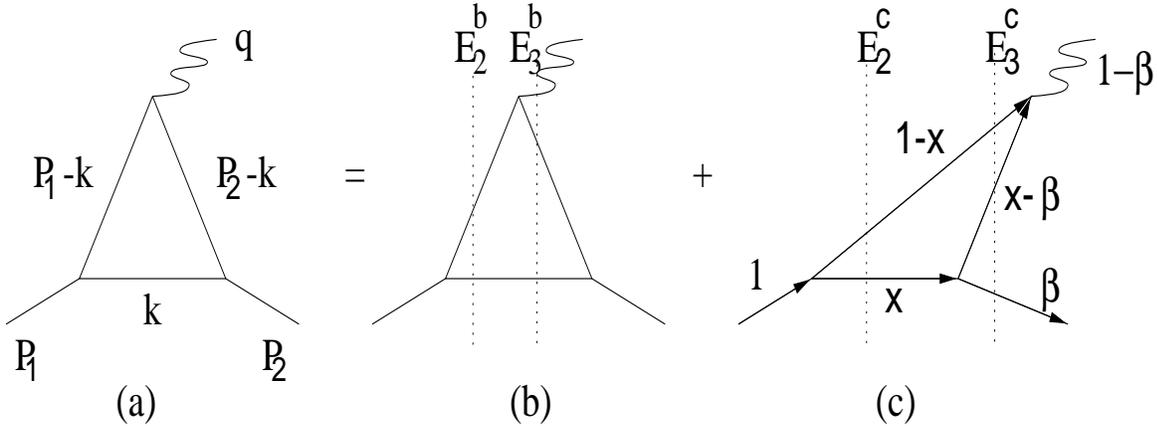,width=6in,height=2.2in}
\caption{Covariant triangle diagram (a) is represented as the sum of
light-front triangle diagram (b) and the light-front pair-creation
diagram (c).}
\end{figure}
As in the case of the timelike form factor in Sec. II, 
the Cauchy integration of $k^{-}$ in Eq.~(\ref{eq:JLF}) has also two  
contributions to the residue calculations, one coming from the interval 
$0<k^{+}<P^{+}_{2}$ (see Fig. 3(b)) and the other from 
$P^{+}_{2}<k^{+}<P^{+}_{1}$ (see Fig. 3(c)). Again, only the
``good"-current $j^{+}(0)$ 
in Eq.~(\ref{eq:JLF}) is used to obtain the contributions from Figs. 3(b) 
and 3(c).  

In the region of $0<k^{+}<P^{+}_{2}$, the residue is at the
pole of $k^{-}=[m_{\bar{Q}}^{2}+k^{2}_{\perp}-i\epsilon]/k^{+}$, which is 
placed in the lower half of complex-$k^{-}$ plane. Thus, the Cauchy 
integration of $j^{+}$ in Eq.~(\ref{eq:JLF}) over $k^{-}$ in this region 
yields
\bea\label{eq:JLFb}
j^{+}_{b}(0)&=& \pi g^2\int^{P^{+}_{2}}_{0}dk^{+}d^{2}k_{\perp}
\frac{(P_{1}+P_{2}-2k)^{+}}{(P_{1}-k)^{+}(P_{1}-k-q)^{+}k^{+}}
\nonumber\\
&&\times 
\frac{1}{ \biggl[P^{-}_{1}- [m_{q}^{2}+k^{2}_{\perp}]/(P^{+}_{1}-k^{+})
- [m_{\bar{Q}}^{2}+k^{2}_{\perp}]/k^{+}\biggr]}\nonumber\\
&&\times
\frac{1}{\biggl[P^{-}_{1}-q^{-} - 
[m_{q}^{2}+(k_{\perp}+q_{\perp})^{2}]/(P^{+}_{1}-k^{+})
- [m_{\bar{Q}}^{2}+k^{2}_{\perp}]/k^{+}\biggr]},
\eea 
where the subscript $b$ in $j^{+}_{b}(0)$ implies the current in Fig. 3(b).

In the region of $P^{+}_{2}<k^{+}<P^{+}_{1}$, the residue 
is at the pole of $k^{-}= P^{-}_{1} - 
[m_{q}^{2}+k^{2}_{\perp}-i\epsilon]/(P_{1}^{+}-k^{+})$, which is
placed in the upper half of complex-$k^{-}$ plane. Thus, the Cauchy
integration of $j^{+}$ in Eq.~(\ref{eq:JLF}) over $k^{-}$ in this region 
becomes  
\bea\label{eq:JLFc} 
j^{+}_{c}(0)&=& \pi g^2\int^{P^{+}_{1}}_{P^{+}_{2}}dk^{+}d^{2}k_{\perp}
\frac{(P_{1}+P_{2}-2k)^{+}}{(P_{1}-k)^{+}(P_{1}-k-q)^{+}k^{+}}
\nonumber\\
&&\times
\frac{1}{\biggl[P^{-}_{1}-[m_{q}^{2}+k_{\perp}^{2}]/(P^{+}_{1}-k^{+})
- [m_{\bar{Q}}^{2}+k^{2}_{\perp}]/k^{+}\biggr]}
\nonumber\\
&&\times \frac{1}{ \biggl[P^{-}_{2}- P^{-}_{1} +
[m_{q}^{2}+k_{\perp}^{2}]/(P^{+}_{1}-k^{+})
 - [m_{q}^{2}+(q_{\perp}+k_{\perp})^{2}]/(P^{+}_{2}-k^{+})\biggr]}.
\eea
As one can see from Eq.~(\ref{eq:JLFc}), the nonvalence contribution   
(Fig. 3(c)) vanishes only in the $q^{+}=0$ frame. 
Note that, for the spin-one system composed of fermions, however,
Equation~(\ref{eq:JLFc}) does not vanish even in $q^+=0$ frame when the 
longitudinal polarization vector is involved as mentioned in the 
introduction.
In the following, we investigate the spacelike form factor 
${\cal F}(q^{2})$ given in Eq.~(\ref{eq:Jspace}) using both $q^{+}\neq 0$ 
and $q^{+}=0$ frames. We then analytically continue to the timelike region
in order to compare the result with the direct calculation of the timelike
form factor $F(q^2)$ presented in the previous section. 
\subsection{The purely longitudinal($q^{+}\neq0$ and $q_{\perp}$=0) frame}
In the purely longitudinal momentum frame $q^{+}\neq 0$, $q_{\perp}=0$, 
and $P_{1\perp}=P_{2\perp}=0$, the momentum transfer $q^{2}=q^{+}q^{-}$ 
can be written in terms of the longitudinal momentum fraction 
$\beta=P^{+}_{2}/P^{+}_{1}= 1-q^{+}/P^{+}_{1}$;  
\bea\label{eq:q2space}
q^{2}=-M^{2}(1-\beta)^{2}/\beta\leq 0,
\eea
where the two solutions of $\beta$ are given by  
\bea\label{eq:beta}
\beta_{\pm}= \biggl(1 -\frac{q^{2}}{2M^{2}}\biggr)\pm
\sqrt{ \biggl(1-\frac{q^{2}}{2M^{2}}\biggr)^{2} -1}.
\eea   
The form factor ${\cal F}(q^{2})$ in Eq.~(\ref{eq:Jspace}) is also 
independent of the subscript sign of $\beta$. 
However, the condition $P_{2}^{+}\leq P^{+}_{1}$ was used in
obtaining Eqs.~(\ref{eq:JLFb}) and~(\ref{eq:JLFc}) and thus here we 
use $\beta=\beta_{-}$ ($0\leq\beta\leq 1$) in spacelike region.  
As shown in Eqs.~(\ref{eq:JLF})-(\ref{eq:JLFc}), the sum of valence 
(Fig. 3(b)) and nonvalence (Fig. 3(c)) diagrams is equivalent to the 
covariant triangle diagram in Fig. 3(a). 

For the analysis of singularity structures, we integrate over 
$k_{\perp}$ and obtain from the valence contribution(Fig. 3(b));
\bea\label{eq:JLFb1} 
j^{+}_{b}(0)= \pi^2 g^2\int^{1}_{0}dx 
\frac{\beta x(1+\beta-2\beta x)}{\tilde{{\cal E}}_{3}^{b}
-\tilde{{\cal E}}_{2}^{b}}
\ln\biggl(\frac{\tilde{{\cal E}}_{2}^{b}}{\tilde{{\cal E}}_{3}^{b}}\biggr),
\eea
where $\tilde{{\cal E}}_{3}^{b}= x(1-x)M^{2} -[xm_{q}^{2} + 
(1-x)m_{\bar{Q}}^{2}]$ 
and $\tilde{{\cal E}}_{2}^{b}= \beta x(1-\beta x)M^{2} 
- [\beta xm_{q}^{2} + (1-\beta x)m_{\bar{Q}}^{2}]$.
It turns out that Eq.~(\ref{eq:JLFb1}) has no singularities because
$\tilde{{\cal E}}_{3}^{b}\neq 0$, $\tilde{{\cal E}}_{2}^{b}\neq 0$,
and $\tilde{{\cal E}}_{3}^{b}-\tilde{{\cal E}}_{2}^{b}\neq 0$
for the entire $q^{2}$ region.
On the other hand, the $k_{\perp}$ integration for the 
current $j^{+}_{c}(0)$ in Eq.~(\ref{eq:JLFc}) yields
\bea\label{eq:JLFc1}
j^{+}_{c}(0)= \pi^2 g^2\int^{1}_{0}d{\cal X}  
\frac{(1-\beta)^{2}{\cal X}(2{\cal X}-1)}{\tilde{{\cal E}}_{2}^{c}
-\tilde{{\cal E}}_{3}^{c}}
\ln\biggl(\frac{\tilde{{\cal E}}_{3}^{c}}{\tilde{{\cal E}}_{2}^{c}}\biggr),
\eea
where $x=1-(1-\beta){\cal X}$, 
$\tilde{{\cal E}}_{3}^{c}=(1-\beta){\cal X}[1-(1-\beta){\cal X}]M^{2}-
[(1-\beta){\cal X}m_{\bar{Q}}^{2} + (1 - (1-\beta){\cal X})m_{q}^{2}]$ and
$\tilde{{\cal E}}_{2}^{c}=-(1-\beta)^{2}{\cal X}(1-{\cal X})M^{2}/\beta 
-m_{q}^{2}$. 
While $\tilde{{\cal E}}_{2}^{c}$ corresponding to the energy denominator
of the two
particle intermediate state does not vanish both in spacelike and timelike  
region, $\tilde{{\cal E}}_{3}^{c}$ of the 
three-particle intermediate state is not zero only for the
spacelike momentum transfer region. For the timelike region,
$\tilde{{\cal E}}_{3}^{c}$ can be zero so that
the singularities start at $q^{2}_{\rm min}=4m^{2}_{q(\bar{Q})}$ for 
$\gamma^{*} q\bar{q}(\gamma^{*} Q\bar{Q})$ vertex.
The singularity structure of 
$\tilde{{\cal E}}_{2}^{c}-\tilde{{\cal E}}_{3}^{c}$ in Eq.~(\ref{eq:JLFc1}) 
is the same as in the case of timelike form factor (Section II), following
the condition of a $q\bar{Q}$ bound state.

The EM form factor ${\cal F}(q^{2})$ in Eq.~(\ref{eq:Jspace}) of a 
$q\bar{Q}$ bound state in $q^{+}\neq0$ frame is then obtained by
\bea\label{eq:FNDY}
{\cal F}(q^{2},q^{+}\neq0)&=&
\frac{\pi^2 g^2}{1+\beta}\int^{1}_{0}dx
\biggl\{\frac{\beta x(1+\beta-2\beta x)}{\tilde{{\cal E}}_{3}^{b}
-\tilde{{\cal E}}_{2}^{b}}
\ln\biggl(\frac{\tilde{{\cal E}}_{2}^{b}}{\tilde{{\cal E}}_{3}^{b}}\biggr)
\nonumber\\
&&+ \frac{(1-\beta)^{2}x(2x-1)}{\tilde{{\cal E}}_{2}^{c}
-\tilde{{\cal E}}_{3}^{c}}
\ln\biggl(\frac{\tilde{{\cal E}}_{3}^{c}}{\tilde{{\cal E}}_{2}^{c}}\biggr)
\biggr\},
\eea
where ${\tilde{{\cal E}}_{2}^{b}},{\tilde{{\cal E}}_{3}^{b}},
{\tilde{{\cal E}}_{2}^{c}}$ and ${\tilde{{\cal E}}_{2}^{c}}$ are defined
in Eqs.~(\ref{eq:JLFb1}) and~(\ref{eq:JLFc1}). 
Here, $\beta$ is a function of $q^2$. According to the analytic 
continuation, the sign of $q^{2}$ in Eq.~(\ref{eq:FNDY}) must be changed 
from $-$ to $+$ for the timelike region.

\subsection{ The Drell-Yan-West ($q^{+}$=0) frame} 
In Drell-Yan-West frame,   
$q^{+}=0$, $q^{2}=-q^{2}_{\perp}$, and $P_{1\perp}=P_{2\perp}=0$, 
the `$+$' component of the current 
has only the valence contribution, i.e., $j^{+}_{b}(0)$ in
Eq.~(\ref{eq:JLFb}). 
The current $j^{+}_{b}(0)$ in $q^{+}=0$ frame is given by 
\bea\label{eq:jbDY}
j^{+}_{b}(0)&=& \pi g^2\int^{1}_{0}dx d^{2}\ell_{\perp}
\frac{2x(1-x)}{({\cal A}-\ell^{2}_{\perp}-\xi^{2})^{2}
-4\xi^{2}\ell^{2}_{\perp}\cos^{2}\phi},
\eea
where $\ell_{\perp}=k_{\perp}+xq_{\perp}/2$, 
${\cal A}=x(1-x)M^{2}-[xm^{2}_{q} + (1-x)m^{2}_{\bar{Q}}]$, and 
$\xi^{2}=x^{2}q^{2}_{\perp}/4$. The angle $\phi$ ($0\leq\phi\leq 2\pi$) is 
defined by ${\bf\ell}_{\perp}\cdot{\bf q}_{\perp}
=|{\bf\ell}_{\perp}||{\bf q}_{\perp}|\cos\phi$. 

Integrating Eq.~(\ref{eq:jbDY}) over $\phi$ and $\ell_{\perp}$, we obtain 
\bea\label{eq:jbDY2}
j^{+}_{b}(0)&=& 
-\frac{8\pi^2 g^2}{\sqrt{q^{2}_{\perp}(4M^{2}+q^{2}_{\perp})}}
\int^{1}_{0}dx\frac{(1-x)}{\sqrt{(x-x_{+})(x-x_{-})}}\nonumber\\
&&\times
\tanh^{-1}\biggl[\sqrt{\frac{q^{2}_{\perp}}{4M^{2}+q^{2}_{\perp}}}
\frac{x}{\sqrt{(x-x_{+})(x-x_{-})}}\biggr],
\eea  
where 
\bea\label{eq:xpm}
x_{\pm}=\frac{2(M^{2}-m_{q}^{2}+m_{\bar{Q}}^{2})}{4M^{2}+q^{2}_{\perp}}\pm
\sqrt{\frac{4(M^{2} - m_{q}^{2} + m_{\bar{Q}}^{2})^{2}}
{(4M^{2}+q^{2}_{\perp})^{2}} -
\frac{4m_{\bar{Q}}^{2}}{(4M^{2}+q^{2}_{\perp})} }.
\eea 
The analytic continuation from spacelike to timelike region
in the $q^{+}=0$ frame requires the change of
$q_{\perp}$ to $iq_{\perp}$ in Eqs.~(\ref{eq:jbDY2}) and~(\ref{eq:xpm}). 
We note from Eqs.~(\ref{eq:jbDY2}) and~(\ref{eq:xpm}) that the result of
the timelike region exhibits the same singularity structure 
as the direct analyses in $q^{+}\neq 0$ 
frame, i.e., Eqs.~(\ref{eq:JbP1}) and~(\ref{eq:JcP1}), even though the 
nonvalence contribution in Fig. 3(c) is absent here.  

After some manipulation, we obtain the EM form factor of a 
$q\bar{Q}$ bound state in the $q^{+}=0$ frame as follows  
\bea\label{eq:FDY}
{\cal F}(q^{2},q^{+}=0)&=&-\frac{4\pi^2 g^2}{\sqrt{q^{2}(q^{2}-4M^{2})}}
\int^{\arcsin(\frac{1-a}{b})}_{\arcsin(-\frac{a}{b})}
d\theta (1-a-b\sin\theta)\nonumber\\
&&\times 
\tanh^{-1}\biggl[\sqrt{\frac{q^{2}}{q^{2}-4M^{2}}} 
\frac{a+b\sin\theta}{ib\cos\theta}\biggr],
\eea  
where $a=(x_{+}+x_{-})/2$, $b=(x_{+}-x_{-})/2$, and  
$q^{2}=-q^{2}_{\perp}$.
While the representations in Eqs.~(\ref{eq:FNDY}) and~(\ref{eq:FDY}) look 
apparently different, the two formulas, 
Eqs.~(\ref{eq:FNDY}) and~(\ref{eq:FDY}), turn
out to be actually identical. As we will show explicitly in the next
section of numerical calculations, all three results of 
Eqs.~(\ref{eq:EMtime}), ~(\ref{eq:FNDY}) and~(\ref{eq:FDY})
indeed coincide exactly in the entire $q^2$ range. 
\section{Numerical Results} 
For our numerical analysis of $\pi$, $K$, and $D$ meson form factors, 
we use the physical meson masses together with the following constituent 
quark and antiquark masses: $m_{u}=m_{d}=0.25$ GeV, $m_{s}= 0.48$ GeV, and 
$m_{c}= 1.8$ GeV\cite{CJ2,CJ1,CJ3}. 
Since our numerical results of the EM form factors obtained from 
Eqs.~(\ref{eq:EMtime}), ~(\ref{eq:FNDY}) and~(\ref{eq:FDY})
turn out to be exactly same with each other for the entire
$q^{2}$ region, only a single line is depicted in Figs. 4, 5 and 6 for the
form factor calculations of
$\pi,K$, and $D$ mesons, respectively.

In should be noted from our constituent masses
that $M^{2}< m_{q}^{2} + m_{\bar{Q}}^{2}$ for $\pi$ and $K$ and
$M^{2}> m_{q}^{2} + m_{\bar{Q}}^{2}$ for $D$ meson cases.
As discussed in Ref.\cite{Gasio} for the analysis of the
one-particle matrix element of a scalar current, 
the sigularity for
$M^{2}> m_{q}^{2} + m_{\bar{Q}}^{2}$ case starts at
\bea\label{eq:qmin}
q^{2}_{\rm min}=\frac{1}{m^{2}_{\bar{Q}(q)}}[ m^{2}_{q(\bar{Q})} -
(M - m_{\bar{Q}(q)})^{2}][ (M + m_{\bar{Q}(q)})^{2}
- m^{2}_{q(\bar{Q})}]
\eea
for $\gamma^{*} q\bar{q}(\gamma^{*} Q\bar{Q})$ vertex,
while the singularity for
$M^{2}< m_{q}^{2} + m_{\bar{Q}}^{2}$ case starts on the positive $q^{2}$-axis
at the threshold point $q^{2}_{\rm min}=4m_{q(\bar{Q})}^{2}$ for 
$\gamma^{*} q\bar{q}(\gamma^{*} Q\bar{Q})$ vertex.
Our numerical results exhibit all of these threshold 
behaviors coming from the normal ($\pi,K$) and anomalous ($D$) cases.  
As a consistency check, we also compare  
our numerical results of the form factor $F(q^{2})= {\rm Re}\;
F(q^{2}) + i\;{\rm Im}\; F(q^{2})$ with the dispersion relations given by
\begin{eqnarray}
{\rm Re}\;F(q^{2})&=&\frac{1}{\pi}P\int^{\infty}_{-\infty}
\frac{{\rm Im}\;F(q'^{2})}{q'^{2}-q^{2}}dq'^{2},\label{eq:Re}\\
{\rm Im}\;F(q^{2})&=&-\frac{1}{\pi}P\int^{\infty}_{-\infty}
\frac{{\rm Re}\;F(q'^{2})}{q'^{2}-q^{2}}dq'^{2},\label{eq:Im}
\end{eqnarray}
where $P$ indicates the Cauchy principal value. 

In Fig. 4(a), we show the EM form factor of the pion for
$-2\;{\rm GeV}^{2}\leq q^{2}\leq 3\;{\rm GeV}^{2}$. 
The imaginary part (the dotted line) 
of the form factor starts at $q^{2}_{\rm min}=4m^{2}_{u(d)}=0.25$ 
GeV, which is consistent with the condition for 
$M^{2}< m_{q}^{2} + m_{\bar{Q}}^{2}$ case. It is interesting to note that  
the square of the total form factor $|F_{\pi}(q^{2})|^{2}$ (thick solid line) 
produces a $\rho$ meson-type peak near $q^{2}\sim M^{2}_{\rho}$. 
However, we do not claim that this model indeed reproduces all the features
of the vector meson dominance (VMD) phenomena because the more realistic
phenomenological models may have to incorporate the more complex mechanism
including the final state interaction.
More investigation along this line is under consideration. Nevertheless,
this simple model is capable of generating the peak and the position of 
peak is quite consistent with the VMD.

In Fig. 4(b), we show the timelike form factor of the pion for the entire
$q^{2}>0$ region and compare the imaginary part of our direct calculations 
(dotted line) obtained from Eqs.~(\ref{eq:EMtime}),~(\ref{eq:FNDY}), 
and~(\ref{eq:FDY}) with the result 
(data of black dots) obtained from the dispersion relations given by 
Eq.~(\ref{eq:Im}). Our direct calculation is in an excellent agreement 
with the solution of the dispersion relations. Our results for the real 
part are also confirmed to be in complete agreement with the dispersion
relations. For high $q^{2}$ region, the
imaginary part of the form factor is dominant over the real part 
(thin solid line).

In Fig. 5(a), we show the kaon form factor for 
$-2\;{\rm GeV}^{2}\leq q^{2}\leq 5\;{\rm GeV}^{2}$. The kaon also has 
the normal singularity. 
However, it has two thresholds for the imaginary parts; one is
$q^{2}_{\rm min}= 4m^{2}_{u}$ and the other is 
$q^{2}_{\rm min}= 4m^{2}_{s}$.
These lead to the humped shape (dotted line) of the 
imaginary part shown in Fig. 5(a). 
While we have in principle two vector-meson-type peaks (i.e.
$\rho$ and $\phi$), one can see in Fig. 5(a) only 
$\phi$ meson-type peak for the timelike kaon EM form factor above
the physical threshold at $q^{2}_{\rm min}=4M^{2}_{K}$. 
We also show in Fig. 5(b) the imaginary part from our direct calculation is 
in an excellent agreement with the result (data of black dots)
from the dispersion relations   
for the entire timelike $q^{2}$ region. Again, the imaginary part is 
predominant for high $q^{2}$ region.

In Fig. 6, we show the $D$ meson form factor for
$-10\;{\rm GeV}^{2}\leq q^{2}\leq 30\;{\rm GeV}^{2}$. 
Unlike the normal threshold of $\pi$ and $K$ form factor calculations,
the $D$ meson form factor shows anomalous thresholds according to 
Eq.~(\ref{eq:qmin}),
i.e., $q^{2}_{\rm min}\sim 0.24$ GeV$^{2}$ (compared to $4m^{2}_{d}=0.25$
GeV$^{2}$ for normal case) and $q^{2}_{\rm min}\sim 12.4$ GeV$^{2}$ 
(compared to $4m^{2}_{c}=12.96$ GeV$^{2}$ for normal case) for the 
$\gamma^{*}-\bar{d}$ and $\gamma^{*}-c$ vertices, respectively. 
Similar to the kaon case in Fig. 5, we also have 
two unphysical peaks, i.e., $\rho$ and $J/\psi(1S)$ meson type peaks due
to $\bar{d}$ and $c$ quarks, respectively. 
However, the timelike form factor of $D$ meson has no pole
structure for the the physical $q^{2}\geq 4M^{2}_{D}$ region.
In all of these figures (Figs. 4-6), the numerical result of Eq.~(\ref{eq:FDY})
obtained from $q^{+}=0$ frame without encountering the nonvalence diagram
coincides exactly with the numerical results of Eqs.~(\ref{eq:EMtime}) 
and (\ref{eq:FNDY}) obtained from $q^{+}\neq 0$ frame.

\section{Conclusion and Discussion} 
While it is not a simple task to explore the timelike region with the
framework at hand, we attempted in this work to provide a clear example
demonstrating that one can compute the timelike
form factor without encountering the nonvalence contributions.
We investigated both spacelike and timelike regions of the meson
EM form factor using an exactly solvable model of $(3+1)$ dimensional 
scalar field theory interacting with gauge fields (see Eq.~(\ref{eq:Lag})).
We calculated the form factor in spacelike region using 
the Drell-Yan-West
($q^+=0$) frame and showed that its analytic continuation to the 
timelike region reproduces exactly the direct result of
timelike form factor obtained in the longitudinal momentum ($q^{+}\neq 0$
and $q_{\perp} = 0$) frame. The analytic
continuation of the result in the $q^+=0$ frame to the timelike
region automatically generates the effect of the nonvalence contributions
to the timelike form factor.
Another interesting result in our model calculations is that
the peaks analogous to the VMD were generated and the position
of peaks were indeed quite consistent with the VMD.
Even though phenomenologically more elaborate analyses including the final 
state interaction may be necessary to reproduce the more realistic feature 
of VMD, our results seem pretty encouraging for further investigations.
Using the dispersion relations, we have also confirmed that
our numerical results of imaginary parts start at 
$q^{2}_{\rm min}=4m_{q(\bar{Q})}$ and
the normal thresholds appear for
$\pi$ and $K$ ($M^{2}<m^{2}_{q} +m^{2}_{\bar{Q}}$) systems 
while the anomalous threshold exists for $D$ ($M^{2}>m^{2}_{q}
+m^{2}_{\bar{Q}}$) system.
However, we note that our form factor calculations are effectively
various ways of evaluating the Feynman perturbation theory triangle
diagrams and thus the analytic properties presented in sections II, III, 
and IV are restricted to one-loop diagrams.

Even though the model (Eq.~(\ref{eq:Lag})) may not be as realistic as one 
may want, we think that it plays the role of some guidance for the more 
realistic phenomenological models. 
For an application of our analytic continuation method to the more 
realistic model, we have considered an immediate extension of bosonic 
contact interaction to the fermionic covariant model with the monopole 
smearing vertices used by de Melo et al.~\cite{Mel1,Mel2} 
and investigated singular free processes in timelike
region such as $0^-(=J^P)\to 0^-$ and $0^-\to 1^-$ semileptonic decays. 
In this application, we found that the covariant solution of the weak 
form factors $f_{+}$ for $0^-\to0^-$ and $V$ and $A_2$ for $0^-\to1^-$ 
processes are exactly equal to our analytic solutions obtained 
from the current $j^+$ in $q^+=0$ frame where only valence diagram is 
needed. On the other hand, the form factors $f_{-}$ for 
$0^-\to0^-$ and $A_0$ and $A_1$ for $0^-\to1^-$ processes turn out to be 
affected by the zero-mode contributions and one has to include the 
zero-mode effects to make our analytic solutions from $q^+=0$ frame 
equal to the covariant solutions (see also~\cite{Ja1}).
This covariant model calculation further verifies the validity of our 
analytic continuation method. We also applied our analytic 
continuation method to the same semileptonic decay processes of
various pseudoscalar and vector mesons using our LFQM~\cite{CJ1,CJ3,CJ4} 
and found a good agreement with the available experimental data as well
as the lattice calculations. 

However, an additional step of identifying the zero-mode contributions is
necessary in handling the zero-mode issue associated with the currents 
$j_{\perp}$ for $0^-\to0^-$ and $j^+$ coupled with longitudinal decay
modes for $0^-\to1^-$ exclusive semileptonic decay processes. 
Also, the singular structures around the pole-regions in the timelike 
elastic form factors needs to be further investigated in the more 
phenomenological model.
Nevertheless, it is encouraging that in this first attempt of calculating
the timelike elastic meson form factor, the analytic
continuation method seems to work as nicely as one may have anticipated.

We would like to thank G.-H.Kim  for assisting the numerical analysis of
dispersion relations.
This work was supported by the U.S. DOE under contracts DE-FG02-96ER40947.
The North Carolina Supercomputing Center and the National Energy Research
Scientific Computer Center are also acknowledged for the grant of 
supercomputer time.

\newpage 
\setcounter{figure}{0}
\renewcommand{\thefigure}{\mbox{4.\alph{figure}}}
\begin{figure}
\centerline{\psfig{figure=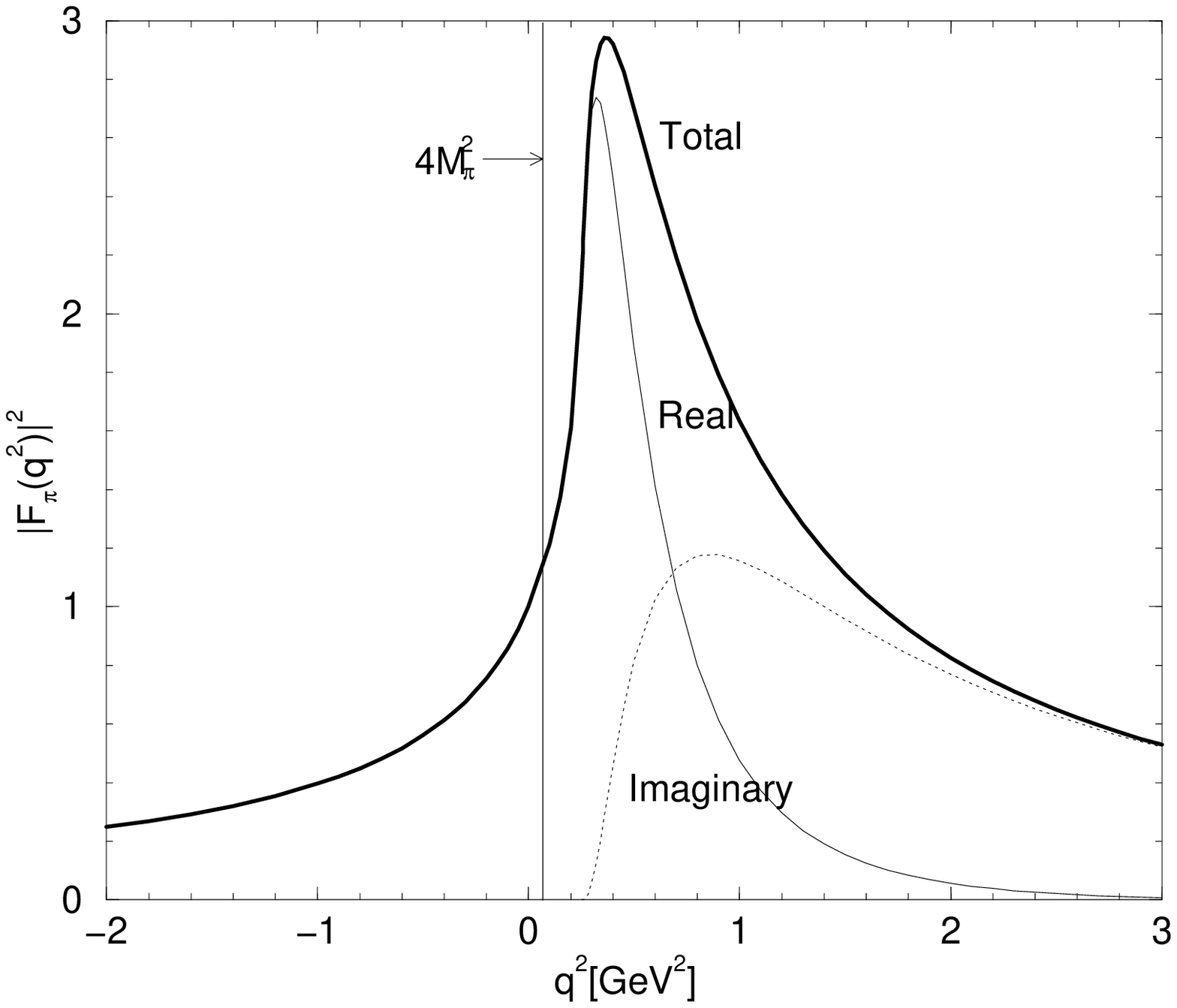,width=3.5in,height=3.5in}}
\caption{The electromagnetic form fator of the pion in $(3+1)$ dimensional
scalar field theory for $-2\leq q^{2}\leq 3$ GeV$^{2}$. The total, 
real, and imaginary parts of $|F_{\pi}(q^{2})|^{2}$ are represented 
by thick solid, solid, and dotted lines, respectively.} 
\end{figure}
\begin{figure}
\centerline{\psfig{figure=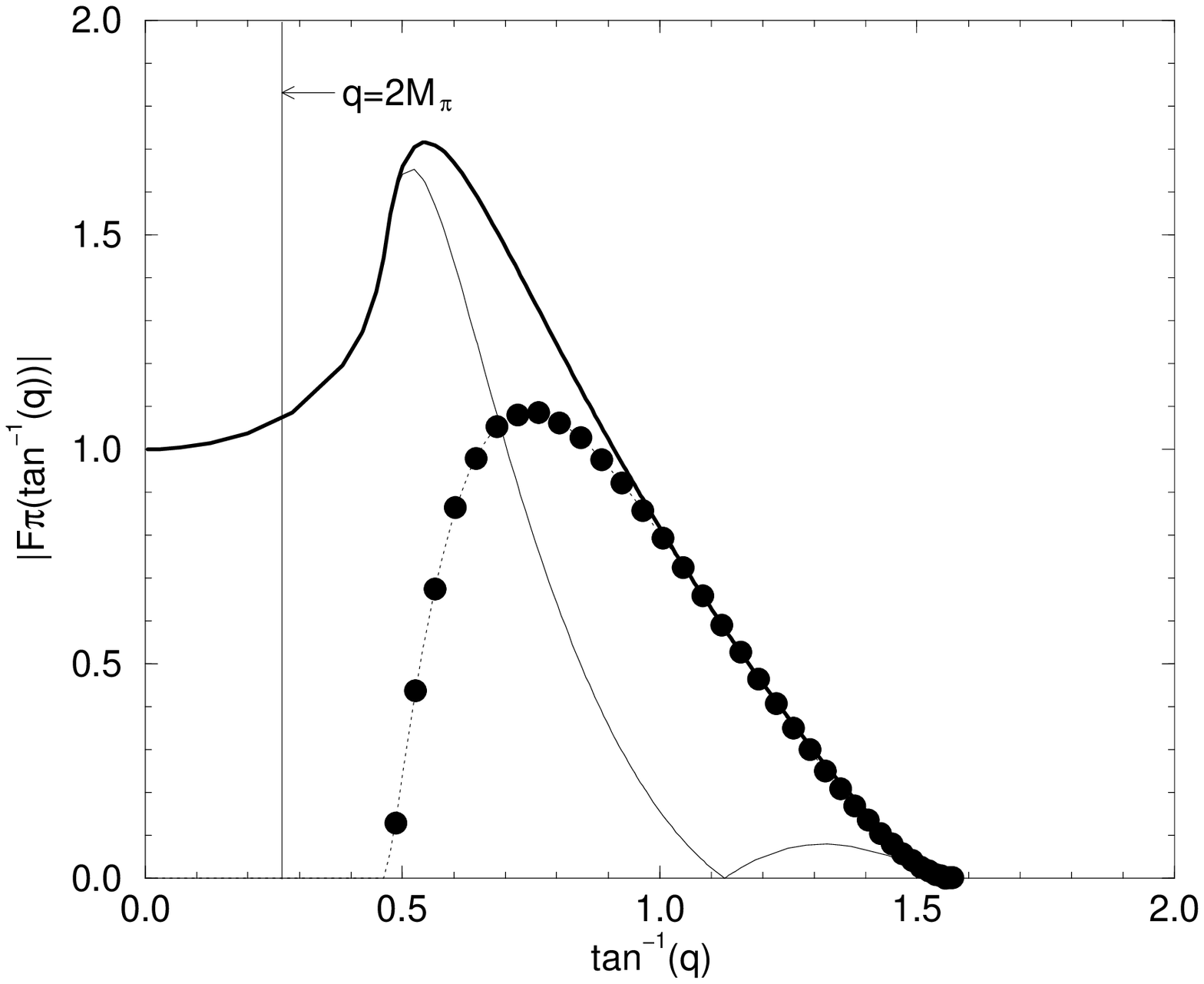,width=3.5in,height=3.5in}}
\caption{The electromagnetic form fator of the pion in $(3+1)$ dimensional
scalar field theory for the entire timelike region compared to the
dispersion relations (data of black dots) given by Eq. (25). 
The same line code as in Fig. 4(a) is used.}
\end{figure}
\setcounter{figure}{0}
\renewcommand{\thefigure}{\mbox{5.\alph{figure}}}
\begin{figure}
\centerline{\psfig{figure=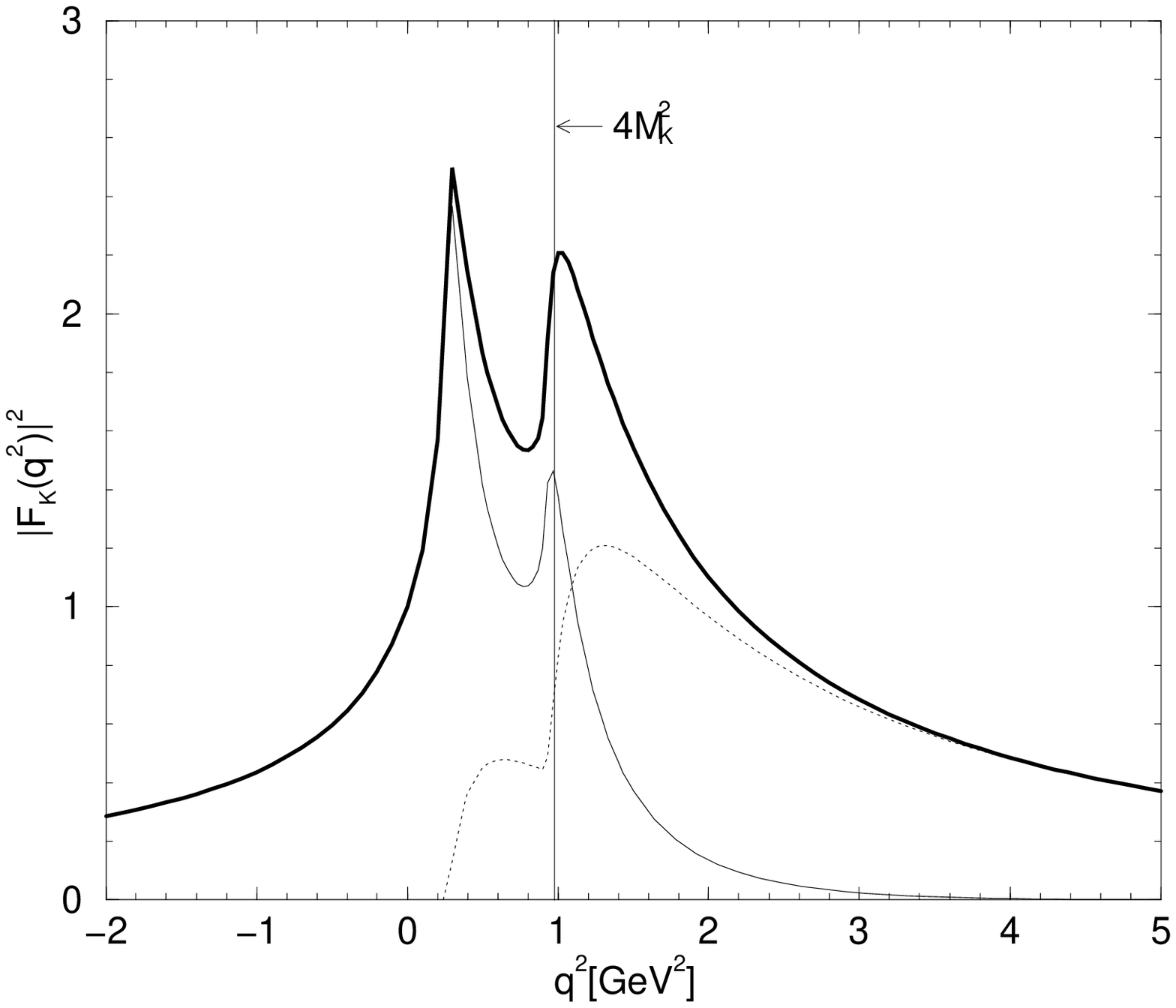,width=3.5in,height=3.5in}}
\caption{The electromagnetic form fator of the kaon in $(3+1)$ dimensional
scalar field theory for $-2\leq q^{2}\leq 5$ GeV$^{2}$. The same line code as 
in Fig. 4(a) is used.}
\end{figure}
\begin{figure}
\centerline{\psfig{figure=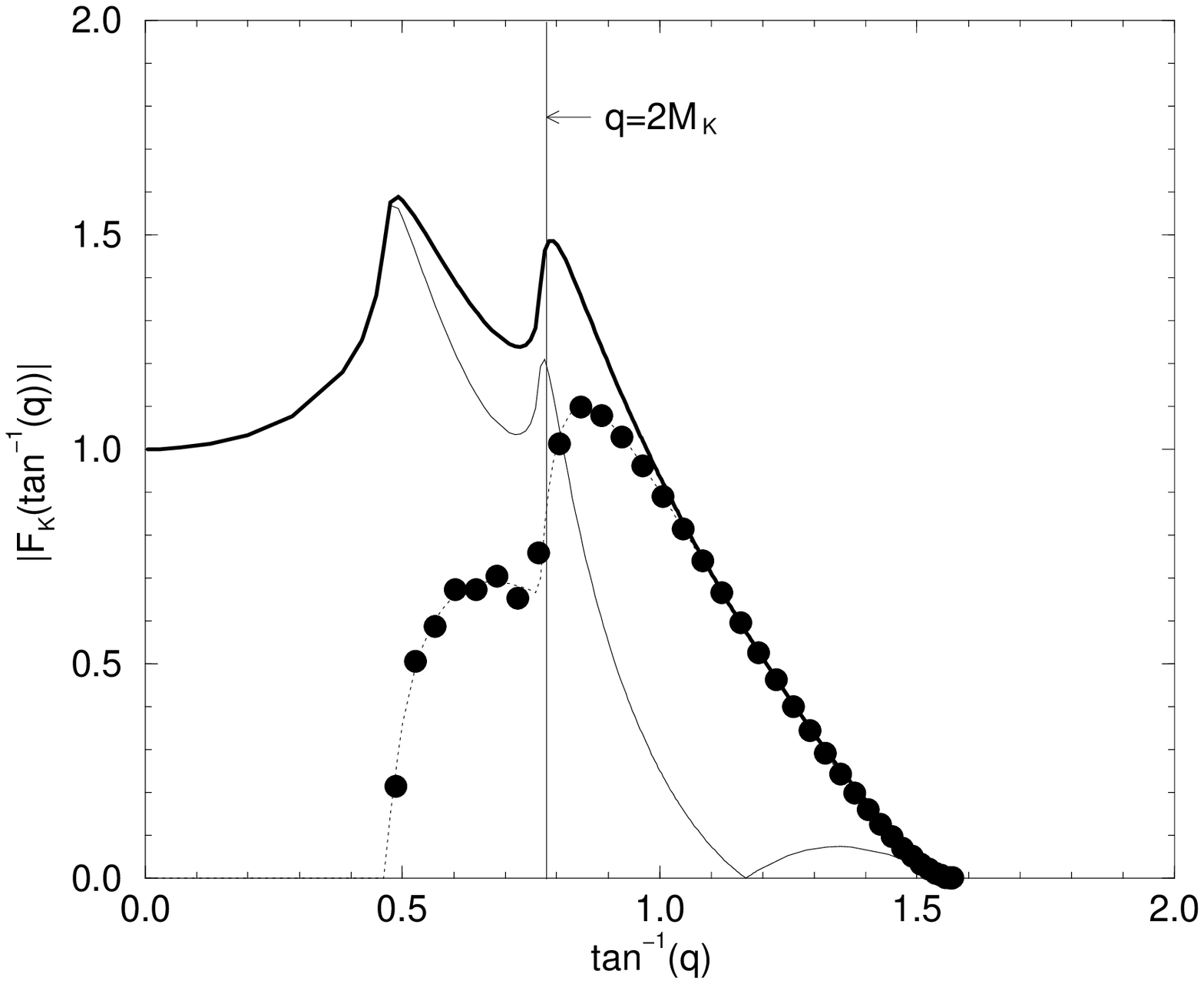,width=3.5in,height=3.5in}}
\caption{The electromagnetic form fator of the kaon in $(3+1)$ dimensional
scalar field theory for the entire timelike region compared to the
dispersion relations (data of black dots) given by Eq. (25). 
The same line code as in Fig. 4(a) is used.}
\end{figure}
\renewcommand{\thefigure}{\mbox{6}}
\begin{figure}
\centerline{\psfig{figure=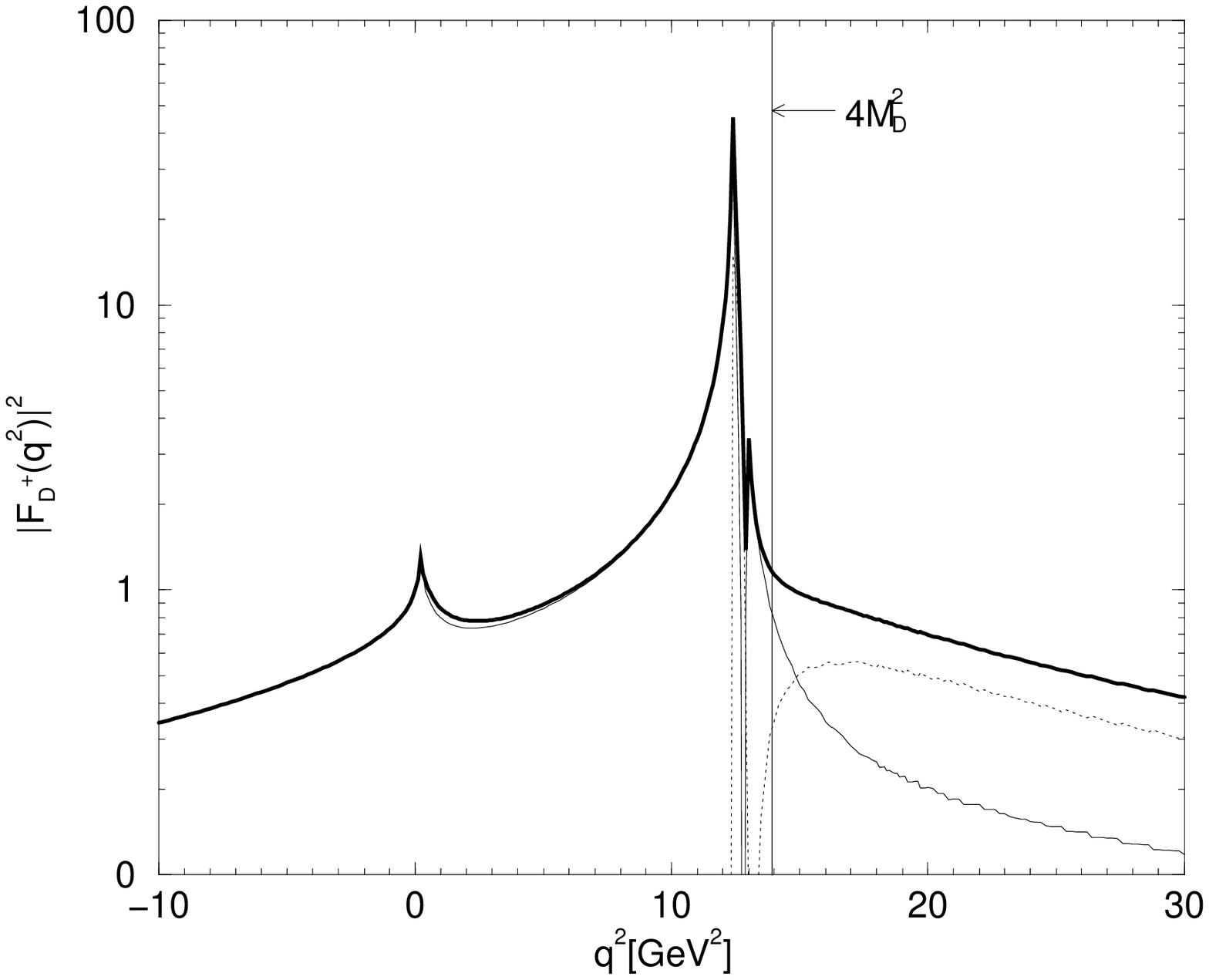,width=3.5in,height=3.5in}}
\caption{The electromagnetic form fator of the $D$ meson in $(3+1)$ dimensional
scalar field theory for $-10\leq q^{2}\leq 30$ GeV$^{2}$. The same line code as
in Fig. 4(a) is used.}
\end{figure}
\end{document}